\documentclass[reprint,aps,prl,showpacs,twocolumn]{revtex4-1}
\usepackage{eurosym}
\usepackage{mathptmx}
\usepackage{epsfig,amsopn}
\usepackage{graphicx}
\usepackage{amsmath,amssymb}
\usepackage{natbib,color}
\usepackage{braket}
\usepackage{float}
\usepackage{enumerate}

\pdfoutput=1
\allowdisplaybreaks[1]

\def\bea{\begin{eqnarray}}
\def\eea{\end{eqnarray}}

\begin{document}

\title{Gumbel Central Limit Theorem for Max-Min and Min-Max}

\author{{\normalsize{}Iddo Eliazar$^{1}$}
{\normalsize{}}}
\email{eliazar@tauex.tau.ac.il}

\author{{\normalsize{}Ralf Metzler$^2$}{\normalsize{}}}\email{rmetzler@uni-potsdam.de}

\author{{\normalsize{}Shlomi Reuveni$^{1},$}
{\normalsize{}}}
\email{shlomire@tauex.tau.ac.il}

\affiliation{\noindent \textit{
$^{1}$School of Chemistry, The Center for Physics and Chemistry of Living Systems, The Raymond and Beverly Sackler Center for Computational Molecular and Materials Science,\\ \& The Mark Ratner Institute for Single Molecule Chemistry, Tel Aviv University, Tel Aviv 6997801, Israel\\
$^{2}$University of Potsdam, Institute of Physics \& Astronomy, 14476 Potsdam,
Germany}}

\begin{abstract}
\noindent{The Max-Min and Min-Max of matrices arise prevalently in
science and engineering. However, in many real-world situations the computation of the Max-Min and Min-Max is challenging as matrices are large and full information about
their entries is lacking. Here we take a statistical-physics approach and establish limit-laws -- akin to the Central Limit
Theorem -- for the Max-Min and Min-Max of large random matrices. The limit-laws intertwine random-matrix theory and extreme-value theory, couple the matrix-dimensions geometrically, and assert that Gumbel statistics emerge irrespective of the matrix-entries' distribution. Due to their generality and universality, as well as their practicality, these novel results are expected to have a host of applications in the physical sciences and beyond.}
\end{abstract}

\maketitle

The Central Limit Theorem (CLT) -- a foundational cornerstone of statistical physics and probability theory -- is of prime importance in science and engineering. The CLT and its generalized version assert that the scaled sum of a large number of independent and identically distributed (IID) random variables is governed, asymptotically, by two limit-law statistics \cite{Fel1,Fel2}: Normal and L{\'{e}}vy-stable. The CLT considers finite-variance IID random variables, and yields Normal statistics. Departing the finite-variance dominion, the generalized CLT imposes sharp tail conditions on the distribution of the IID random variables \cite{BGT}, and yields both Normal and L{\'{e}}vy-stable statistics.

Extreme-value theory \cite{Cas,RT} is applied whenever extreme behavior -- rather than average behavior -- is of relevance; e.g. the prediction of rare events, and the safe design of critical systems such as dams, bridges, and power grids. Extreme-value theory shifts the focus from sums to extrema, i.e. maxima and minima. The Fisher-Tippett-Gnedenko (FTG) theorem is the extreme-value counterpart of the above CLTs. This theorem asserts that the scaled extrema of a large number of IID random variables are governed, asymptotically, by three limit-law statistics \cite{Gne,Gum}: Weibull, Frechet, and Gumbel. As in the case of the generalized CLT, the FTG theorem imposes sharp tail conditions on the distribution of the IID random variables \cite{BGT}. 

The limit-law statistics of the CLTs and the FTG theorem play key roles in physics, e.g. in \cite{Cha,BG,SZK,SZF,Tsa,MK,Red,KS2,CMK,BMS,KS1,AKN,SBA} and in \cite{BM,FB,Majumdar1,Majumdar2,Chupeau,STZ,TVCZ}, respectively. Underlying these theorems is a common bedrock: a random-vector setting, with the IID random variables being the vector entries. Elevating from one-dimensional to two-dimensional arrays, we arrive at a random-matrix setting: matrices whose entries are IID random variables. Random matrices also play key roles in physics \cite{Potters1,Potters2}, and much effort has been directed to the extreme-value analysis of their eigenvalues spectra \cite{Majumdar3,Vergassola}. Here we focus on a different extreme-value analysis of random matrices: their Max-Min and Min-Max (see Fig. 1 for the Max-Min).

\begin{figure}[t]
\includegraphics[width=8cm]{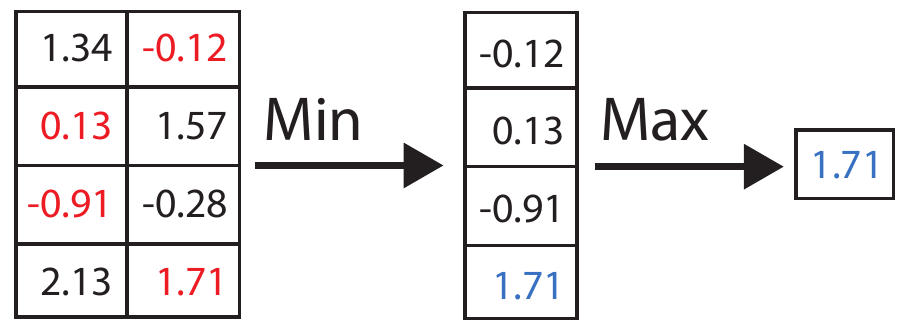}
\caption{The Max-Min of a matrix is obtained by first taking the
minimal entry of each row (depicted red), and then taking the maximum of
these minimal entries (depicted blue).}
\label{fig1}
\end{figure}

The Max-Min and Min-Max arise prevalently in science and engineering. Perhaps the best known example is in game theory \cite{MZS}, a field which drew considerable attention from physicists \cite{GTPHYS1,GTPHYS2,GTPHYS3,GTPHYS4,GTPHYS5,GTPHYS6,GTPHYS7}. There, a player seeks a strategy that will maximize gain, or minimize loss, in the worst-case scenario. The player has a payoff matrix which specifies the
gain/loss for each strategy taken vs. each scenario encountered; the player calculates the Max-Min in the case of
gains, and the Min-Max in the case of losses. However, in real-life situations the payoff matrix is often large and full
information about its entries is lacking. In turn, such situations call for a modeling approach employing large random matrices. 

The Max-Min and Min-Max of large random matrices were investigated in mathematics \cite{CT}, and in reliability engineering \cite{Kolo1,Kolo2,Kolo,RC}. In the pioneering work \cite{CT}, Chernoff and Teicher established that the scaled Max-Min and Min-Max are governed, asymptotically, by the FTG statistics: Weibull, Frechet, and Gumbel. In subsequent works \cite{Kolo1,Kolo2,Kolo}, Kolowrocki further advanced the topic in the context of (so called) series-parallel and parallel-series systems. In a more recent work \cite{RC}, Reis and Castro obtained Gumbel limit-law statistics for the Max-Min via an iterative application of the FTG theorem: first to the minimum of each and every matrix row, and then to the maximum of the rows’ minima. 

The results in \cite{CT,Kolo1,Kolo2,Kolo,RC} are notable and inspiring mathematical theorems. However, from a practical perspective the application of these results is extremely challenging, even on a case by case basis. More importantly, the results in \cite{CT,Kolo1,Kolo2,Kolo,RC} do not provide a clear-cut answer to the following focal question: is there a ``Central Limit Theorem'' for the Max-Min and Min-Max of random matrices?

The CLTs and the FTG theorem stand on two pillars: domain of attraction and scaling scheme. The domain of attraction of the CLT is wide (encompassing all finite-variance distributions), and its scaling scheme is simple; the application of the CLT is thus straightforward, and its use is omnipresent. For the generalized CLT and the FTG theorem matters are more intricate: the domains of attraction are narrow (characterized by the sharp conditions imposed on the distributions' tails \cite{BGT}), and the scaling schemes are elaborate (they need to be carefully custom-tailored per each admissible distribution \cite{BGT}). Elevating from a random-vector setting to a random-matrix setting adds a third pillar to the two above: the asymptotic coupling between the matrix dimensions (as these are taken to infinity). In \cite{CT,Kolo1,Kolo2,Kolo,RC} the intricacy of all three pillars is prohibitively high. Consequently, the are no available Max-Min and Min-Max limit-laws with the following features: wide domain of attraction, simple scaling scheme, and simple asymptotic coupling.

\begin{figure}[t]
\includegraphics[width=8cm]{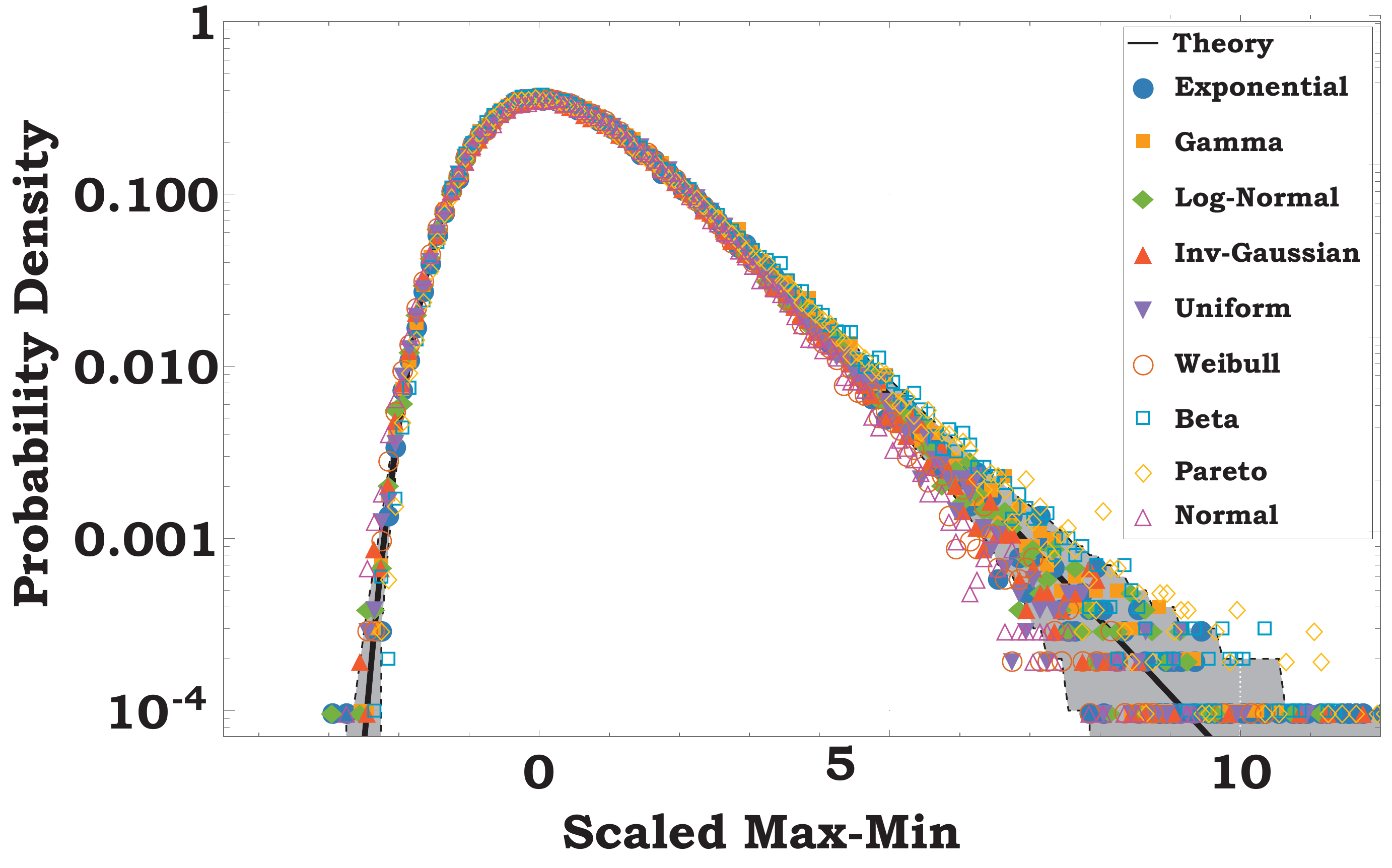}
\caption{Gumbel limit-law statistics for the scaled Max-Min of 
large random matrices. Universality is demonstrated by data collapse for nine
different distributions from which the IID matrix entries are drawn: the
colored symbols depict the simulated data; the solid black line is the
probability density of the predicted Gumbel statistics, with its 95\%
confidence interval shaded in grey.}
\label{fig2}
\end{figure}

Here we present ``Central Limit Theorem'' results for the Max-Min and Min-Max of large non-square random matrices. Circumventing the use of the FTG theorem altogether, the results are based on novel Poisson-process limit-laws \cite{GumP}. The results assert that the scaled Max-Min and Min-Max are governed, asymptotically, by Gumbel statistics. The results' domain of attraction is vast, encompassing all distributions with a density. The results' scaling schemes are similar to that of the CLT, and their asymptotic couplings are geometric. The novel results established here are thus highly practical and applicable (see Fig. 2 for the Max-Min result).

Written for a general physics readership, this rapid communication offers a concise brief of the novel results and their implementation; for a comprehensive exposition, including detailed proofs, see \cite{GumP}. The brief is organized as follows: we begin with an underlying setting, present Gumbel approximations for the Max-Min and Min-Max, and describe the implementation of these approximations; then, we present the Gumbel limit-laws (that yield the Gumbel approximations), discuss these limit-laws, and conclude with an outlook. 

\textit{Setting.}---Consider a random matrix with IID entries:%
\begin{equation}
\mathbf{M}=\left( 
\begin{array}{ccc}
X_{1,1} & \cdots & X_{1,n} \\ 
\vdots & \ddots & \vdots \\ 
X_{m,1} & \cdots & X_{m,n}%
\end{array}%
\right) \text{ .}  \label{00}
\end{equation}%
Namely, the matrix is of dimensions $m\times n$, with rows labeled $%
i=1,\cdots ,m$, and columns labeled $j=1,\cdots ,n$. The matrix entries are
IID copies of a generic real-valued random variable $X$, with probability
density $f\left( x\right) $ ($-\infty <x<\infty $). In what follows we
denote by $F\left( x\right) =\Pr \left( X\leq x\right) $ ($-\infty <x<\infty 
$) the corresponding distribution function, and by $\bar{F}\left( x\right)
=\Pr \left( X>x\right) $ ($-\infty <x<\infty $) the corresponding survival
function.

We set the focus on the Max-Min and Min-Max of the
random matrix $\mathbf{M}$. Denoting by $\wedge _{i}=\min \left\{
X_{i,1},\cdots ,X_{i,n}\right\} $ the minimum over the entries of row $i$,
the Max-Min is the maximum over the rows' minima: 
\begin{equation}
\wedge _{\max }=\max \left\{ \wedge _{1},\cdots,\wedge _{m}\right\} \text{ .}
\label{01}
\end{equation}%
Similarly, denoting by $\vee _{j}=\max \left\{ X_{1,j},\cdots
,X_{m,j}\right\} $ the maximum over the entries of column $j$, the Min-Max
is the minimum over the columns' maxima: 
\begin{equation}
\vee _{\min }=\min \left\{ \vee _{1},\cdots ,\vee _{n}\right\} \text{ .}
\label{02}
\end{equation}

To illustrate the setting, consider the aforementioned game-theory example. If the matrix $\mathbf{M}$ manifests gains then: the rows
represent the player's strategies; the columns represent the scenarios the
player is facing; $X_{i,j}$ is the player's gain when taking strategy $i$
and encountering scenario $j$; and $\wedge _{\max }$ is the player's Max-Min
gain. If the matrix $\mathbf{M}$ manifests losses then the roles of its rows
and columns are transposed, $X_{i,j}$ is the player's loss when encountering scenario $i$ and taking strategy $j$, and $%
\vee _{\min }$ is the player's Min-Max loss. 

From Eqs. (\ref{01}) and (\ref{02}) it follows that the distribution/survival functions of the Max-Min and Min-Max are given, respectively, by $\Pr\left(\wedge_{\max}\leq x\right)=[1-\bar{F}\left( x\right)^n]^m$ and by $\Pr\left(\vee_{\min} > x\right)=[1-F\left( x\right)^m]^n$. In the results to be presented here we scale the Max-Min and Min-Max appropriately, and establish their convergence (in law) to universal Gumbel statistics. In what follows $Z$ denotes a `standard' Gumbel random variable, and $G(x)$ denotes the corresponding Gumbel distribution function \cite{Gum}: 
\begin{equation}
\Pr \left( Z\leq x\right)=G(x)=\exp \left[ -\exp \left( -x\right) \right]   \label{10}
\end{equation}%
($-\infty <x<\infty $). 

Our results involve an `anchor' $x_{\ast }$ -- an arbitrary value that can be realized by the generic random variable $X$. Specifically, the anchor meets two requirements: (i) $0<f\left( x_{\ast }\right) <\infty $%
; and (ii) $0<F\left( x_{\ast }\right) <1$, which is equivalent to $0<\bar{F}%
\left( x_{\ast }\right) <1$. For example, with regard to three of the distributions appearing in Fig. 2, the admissible values of the anchor are: $-\infty
<x_{\ast }<\infty $ for the Normal;  $0<x_{\ast}<\infty $ for the Gamma; and $0<x_{\ast }<1$ for the Beta. 

\textit{Approximations.}---We present Gumbel approximations for the Max-Min $\wedge
_{\max }$ and the Min-Max $\vee _{\min }$ of a large random matrix $\mathbf{M}$ with dimensions $m\gg 1$ and $n\gg 1$. The approximations are based on couplings between the matrix dimensions $m$ and $n$, and the anchor $x_{\ast }$. As we shall show hereinafter, these couplings are always implementable: given two of the triplet $\{m,n,x_{\ast }\}$ we can always set the third to satisfy the couplings. Also, in the approximations $Z$ is the `standard' Gumbel random variable of Eq. (\ref{10}). 

Consider the coupling $m\cdot \bar{F}\left( x_{\ast }\right) ^{n} \simeq 1$; then, the Max-Min admits the approximation
\begin{equation}
\wedge _{\max }\simeq Z_{\max }:=x_{\ast }+\frac{1}{n}\cdot \frac{1}{\alpha }%
Z{\ ,}  \label{21}
\end{equation}%
where $\alpha =f\left( x_{\ast }\right) /\bar{F}\left( x_{\ast }\right) $. Similarly, consider the coupling $n\cdot F\left(x_{\ast }\right) ^{m} \simeq 1$; then, the Min-Max admits the approximation
\begin{equation}
\vee _{\min }\simeq Z_{\min }:=x_{\ast }-\frac{1}{m}\cdot \frac{1}{\beta }Z%
\text{ ,}  \label{22}
\end{equation}%
where $\beta =f\left( x_{\ast }\right) /F\left( x_{\ast }\right) $. 

Eqs. (\ref{21}) and (\ref{22}) imply that: the deterministic approximation of the Max-Min $\wedge_{\max }$ and the Min-Max $\vee _{\min }$ is the anchor $x_{\ast }$; the magnitude of the random fluctuations about $x_{\ast }$ is $1/(n \alpha)$ for the Max-Min, and is $1/(m \beta)$ for the Min-Max; and the statistics of the random fluctuations about $x_{\ast }$ are Gumbel. Key statistical features of the Gumbel approximations $Z_{\max }$ of Eq. (\ref{21}) and $Z_{\min }$ of Eq. (\ref{22}) are detailed in Table 1: modes, medians, means, and standard deviations. The probability densities of the Gumbel approximations $Z_{\max }$ and $Z_{\min }$ have a unimodal shape: monotone increasing below $x_{\ast }$, and monotone decreasing above $x_{\ast }$.

\textit{Implementation.}---There are two ways of implementing the Gumbel approximations, which we now describe. Both ways exploit the couplings underpinning the approximations.  

The first way applies when the matrix dimensions $m$ and $n$ are given; in this case the dimensions determine the anchor $x_{\ast }$. Specifically, for matrix $\mathbf{M}$ with dimensions $m>n\gg 1$ the approximation of Eq. (\ref{21}) holds with anchor $x_{\ast }=\bar{F}^{-1}[\left(1/m\right) ^{1/n}]$. Similarly, for matrix $\mathbf{M}$ with dimensions $n>m\gg 1$ the approximation of Eq. (\ref{22}) holds with anchor $x_{\ast }= F^{-1}[\left(1/n\right) ^{1/m}]$.

The second way applies when the anchor $x_{\ast }$ is given; in this case the matrix dimensions $m$ and $n$ should be set accordingly. Specifically, for the Max-Min setting $n\gg 1$ and $m \simeq 1 / \bar{F}\left( x_{\ast }\right) ^{n}$ yields the approximation of Eq. (\ref{21}). And, for the Min-Max setting $m\gg 1$ and $n \simeq 1/ F\left(x_{\ast }\right) ^{m} $ yields the approximation of Eq. (\ref{22}). In this way the magnitudes of the random fluctuations about the anchor $x_{\ast }$ are: of the order $O(1/n)$ in the approximation of Eq. (\ref{21}), and of the order $O(1/m)$ in the approximation of Eq. (\ref{22}).

The first way is a `scientific tool': given a matrix $\mathbf{M}$, it provides us with approximations for the Max-Min and Min-Max. The second way is an `engineering tool': given a `target' anchor $x_{\ast }$, it tells us how to design the matrix $\mathbf{M}$ so that $x_{\ast }$ will be the deterministic approximation of the Max-Min and Min-Max; moreover, we can design the magnitudes of the random fluctuations about $x_{\ast }$ to be as small as we wish \cite{GumP}. 

\begin{table}[t!]
\centering
\begin{tabular}{|l|l|l|}
\hline
& $%
\begin{array}{c}
\text Z_{\max } \\ 
\end{array}%
$ & $%
\begin{array}{c}
\text Z_{\min } \\ 
\end{array}%
$ \\ \hline
$%
\begin{array}{c}
\text{Mode} \\ 
\end{array}%
$ & $x_{\ast }$ & $x_{\ast }$ \\ \hline
$%
\begin{array}{c}
\text{Median} \\ 
\end{array}%
$ & $x_{\ast }-\frac{\ln \left[ \ln \left( 2\right) \right] }{\alpha }\cdot 
\frac{1}{n}$ & $x_{\ast }+\frac{\ln \left[ \ln \left( 2\right) \right] }{%
\beta }\cdot \frac{1}{m}$ \\ \hline
$%
\begin{array}{c}
\text{Mean} \\ 
\end{array}%
$ & $x_{\ast }+\frac{\gamma }{\alpha }\cdot \frac{1}{n}$ & $x_{\ast }-\frac{%
\gamma }{\beta }\cdot \frac{1}{m}$ \\ \hline
$%
\begin{array}{c}
\text{SD} \\ 
\end{array}%
$ & $\frac{\pi }{\sqrt{6}\alpha }\cdot \frac{1}{n}$ & $\frac{\pi }{\sqrt{6}%
\beta }\cdot \frac{1}{m}$ \\ \hline
\end{tabular}%
\caption{Key statistical features of the Gumbel approximations $Z_{\max }$ of Eq. (\ref{21}) and $Z_{\min }$ of Eq. (\ref{22}): mode, median, mean, and standard deviation (SD); in the row for the mean, $\protect\gamma =0.577\cdots$ is the Euler-Mascheroni constant.}
\label{table:1}
\end{table}

\textit{Limit-laws.}---The Gumbel approximations of Eqs. (\ref{21}) and (\ref{22}) emanate from corresponding Gumbel limit-laws which we now present. In the limit-laws we fix the anchor $x_{\ast }$, and then grow the matrix dimensions infinitely large: $m,n\rightarrow \infty $. Also, in the limit-laws $G(x)$ is the `standard' Gumbel distribution function of Eq. (\ref{10}). 

Grow the matrix dimensions via the coupled limit $\lim_{m,n\rightarrow \infty}m\cdot \bar{F}\left( x_{\ast }\right) ^{n}=1$; then, the Max-Min limit-law is
\begin{equation}
\lim_{m,n\rightarrow \infty}\Pr \left[ \alpha n \left( \ \wedge _{\max
}-x_{\ast }\right) \leq x\right] =G\left( x\right)  \label{11}
\end{equation}%
($-\infty <x<\infty $), where $\alpha =f\left( x_{\ast }\right) /\bar{F}%
\left( x_{\ast }\right) $ as above. Similarly, grow the matrix dimensions via the coupled limit $\lim_{m,n\rightarrow \infty}n\cdot F\left(x_{\ast }\right) ^{m}=1$; then, the Min-Max limit-law is 
\begin{equation}
\lim_{m,n\rightarrow \infty}\Pr \left[ \beta m \left( \ x_{\ast }-\vee
_{\min }\right) \leq x\right] =G\left( x\right)  \label{12}
\end{equation}%
($-\infty <x<\infty $), where $\beta =f\left( x_{\ast }\right) /F\left(
x_{\ast }\right) $ as above. 

\begin{figure*}[t]
\includegraphics[width=17cm]{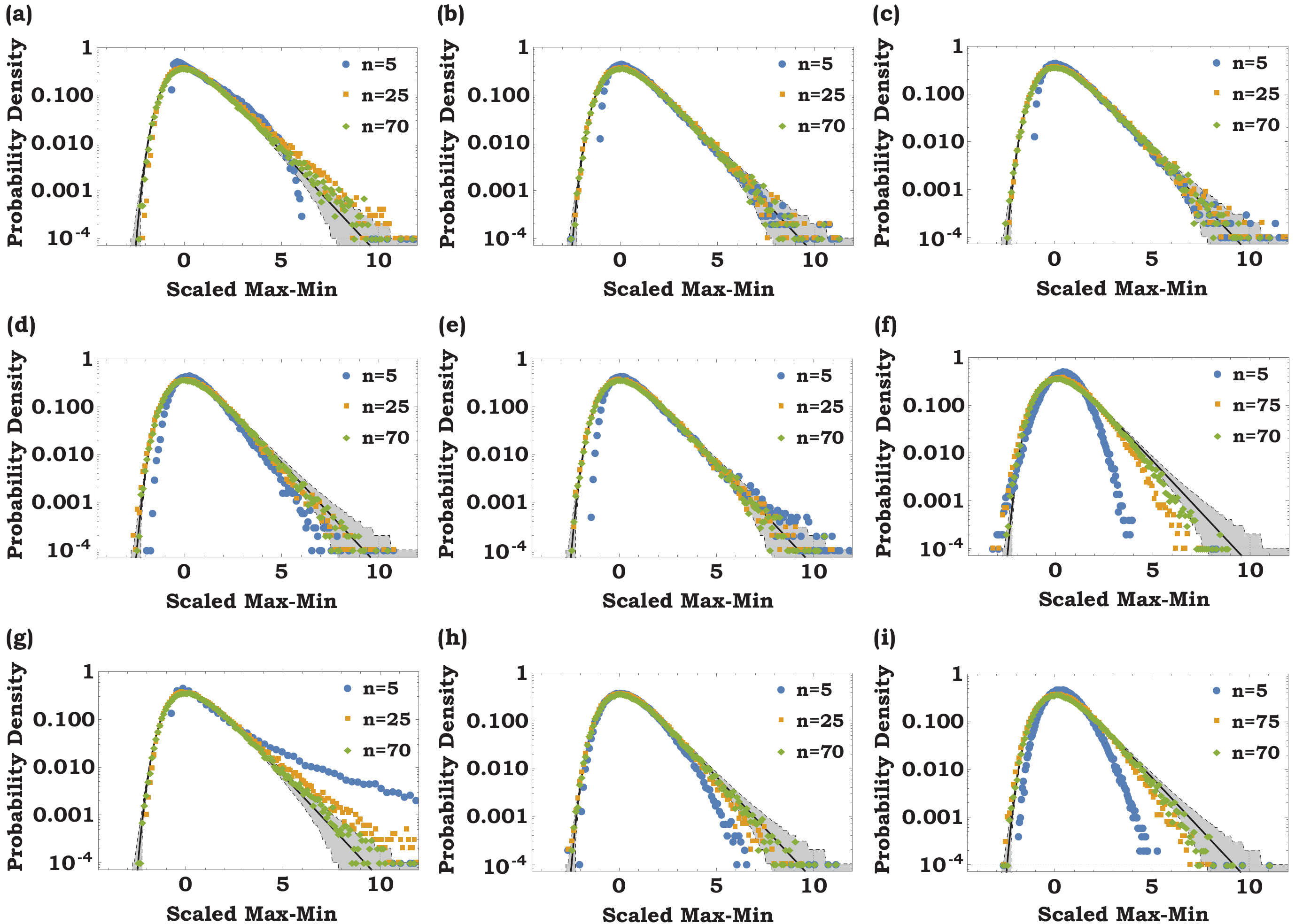}
\caption{The Gumbel limit-law of Eq. (7) is tested for nine different distributions from which the IID matrix entries are drawn: (a) Beta; (b) Exponential; (c) Gamma; (d) Inverse Gaussian; (e) Log-Normal; (f) Normal; (g) Pareto; (h) Uniform; and (i) Weibull. The statistics of the scaled Max-Min $\protect\alpha \cdot n \left(\ \wedge _{\max }-x_{\ast }\right) $, with anchor $x_{\ast }=\bar{F}^{-1} (0.8)$, were simulated by sampling $10^5$ random matrices with the following dimensions: $n=5,25,70$ rows and $%
m \simeq 1.25^n$ columns. In all cases, the convergence of the simulations (colored symbols) to the probability density of the standard Gumbel law (solid black line, with its 95\% confidence interval shaded in grey) is evident.}
\label{fig3}
\end{figure*}

Equations (\ref{11}) and (\ref{12}) imply that the scaled Max-Min $\alpha  n\left( \ \wedge _{\max }-x_{\ast
}\right) $ and the scaled Min-Max $\beta m\left( \ x_{\ast }-\vee _{\min }\right) $ converge -- in law, as $m,n\rightarrow \infty$ -- to a `standard' Gumbel random variable $Z$ (recall Eq. (\ref{10})). Hence, the limit-laws of Eqs. (\ref{11}) and (\ref{12}) yield, respectively, the approximations of Eqs. (\ref{21}) and (\ref{22}). The Gumbel limit-law of Eq. (\ref{11}) is tested for nine different distributions from which the IID matrix entries are drawn (Fig. 3); note that convergence is evident already for moderate values of the dimension $n$. The data collapse demonstrated in Fig. 2 corresponds to the nine distributions of Fig. 3 with dimension $n=70$.

The Gumbel limit-laws of Eqs. (\ref{11}) and (\ref{12}) stem from `bedrock' Poisson-process limit-laws. Underlying the Max-Min $\wedge _{\max }$
is the ensemble of the rows' minima $\left\{ \wedge _{1},\cdots ,\wedge
_{m}\right\} $, and underlying the Min-Max $\vee _{\min }$ is the ensemble
of the columns' maxima $\left\{ \vee _{1},\cdots ,\vee _{n}\right\} $. In 
\cite{GumP} it is established that appropriately scaled versions of these
ensembles converge -- in law, as $m,n\rightarrow \infty $ -- to a Poisson process that is characterized
by the following exponential intensity function: $\lambda (x)=\exp(-x)$ ($-\infty <x<\infty $). For the points of
this Poisson process one can observe that: the maximal point is no
larger than a real threshold $x$ if and only if there are no points above this
threshold -- an event whose probability is $\exp \left[ -\int_{x}^{\infty
}\lambda (x')dx'\right] =G(x)$ \cite{Kingman}. Hence, the distribution
function of the maximal point is $G(x)$ -- which is the term that appears on the right-hand sides of Eqs. (\ref{11}) and (\ref{12}) \cite{GumP}.

\textit{Discussion.}---The limit-laws of Eqs. (\ref{11}) and (\ref{12}) are highly invariant
with respect to the IID entries of the random matrix $\mathbf{M}$. Indeed,
contrary to the CLT -- no moment conditions are imposed on the
entries' distribution. And, contrary to the generalized CLT and to the FTG
theorem -- no tail conditions are imposed on the entries'
distribution. The Gumbel limit-laws merely require that the entries'
distribution has a density. In practice, this smoothness
condition is widely satisfied.


The  Gumbel limit-laws of Eqs. (\ref{11}) and (\ref{12}) involve simple scaling schemes. To appreciate their simplicity, we compare these schemes to that of the CLT. Consider $A_k$ to be the average of $k$ IID random variables with common mean $ \mu $ and standard deviation $ \sigma $. The CLT asserts that the scaled average $\sigma^{-1}\sqrt{k}(A_k - \mu) $ converges -- in law, as $k\rightarrow \infty $ -- to a `standard' Normal random variable (i.e. with a zero mean and a unit standard deviation). The scaled Max-Min $\alpha n\left( \ \wedge _{\max }-x_{\ast}\right) $ of Eq. (\ref{11}) and the scaled Min-Max $\beta m\left( \ x_{\ast }-\vee _{\min }\right) $ of Eq. (\ref{12}) are similar, in form, to the scaled average $ \sigma^{-1}\sqrt{k} (A_k - \mu) $. Specifically: the anchor $x_{\ast }$ is the counterpart of the mean $ \mu $; and the scale terms $\alpha n $ and $\beta m $ are the counterparts of the scale term $\sigma^{-1}\sqrt{k}$. Consequently, the scaling schemes of the limit-laws of Eqs. (\ref{11}) and (\ref{12}) are as simple and straightforward as that of the CLT. 

There are numerously many ways of setting the scaling schemes of the generalized CLT and of the FTG theorem, and each such way corresponds to specific distributions of the underlying IID random variables. On the other hand, as detailed above, the scaling scheme of the CLT is set in a particular way. This special CLT scaling scheme is universal in the following sense: it yields Normal limit-law statistics for all finite-variance distributions.

Addressing limit-laws for the Max-Min and Min-Max of random matrices \cite{CT,Kolo1,Kolo2,Kolo,RC}: there are numerously many ways of setting the scaling schemes; and there are also numerously many ways of asymptotically coupling the matrix dimensions, $m$ and $n$, when growing them infinitely large ($m,n\rightarrow \infty$). Similarly to the CLT, the Gumbel limit-laws of Eqs. (\ref{11}) and (\ref{12}) employ particular scaling schemes, as well as particular asymptotic couplings. In turn, as for the CLT, these special scaling schemes and asymptotic couplings are universal in the following sense: they yield Gumbel limit-law statistics for all distributions with a density.

The particular asymptotic couplings employed here are geometric, and they are parameterized by the anchor $x_{\ast }$. Specifically, the geometric asymptotic couplings are given by: $\lim_{m,n\rightarrow \infty}m\cdot \bar{F}\left( x_{\ast }\right) ^{n}=1$ for the Gumbel limit-law of Eq. (\ref{11}), and $\lim_{m,n\rightarrow \infty}n\cdot F\left(x_{\ast }\right) ^{m}=1$ for the Gumbel limit-law of Eq. (\ref{12}). The couplings' parameterization is a degree-of-freedom that facilitates tunability. Indeed, the anchor $x_{\ast }$ -- which is the counterpart of the mean $ \mu $ in the CLT -- can be tuned as we wish within its admissible values.

\textit{Outlook.}---It has long been observed that seemingly identical pieces of matter happen to fail stochastically at different times and under different loads. Consequently, one of the major original drivers for the development of extreme-value theory came from materials science -- where statistical predictions for mechanical strength and fracture formation are of prime importance \cite{Weibull1,Weibull2}. The ``weakest link hypothesis'' is foundational in materials science \cite{STZ,TVCZ}. This hypothesis suggests that various mechanical systems can be modeled as having a chain-like structure -- thus implying that such a system is only as strong as its weakest link. The ``weakest link hypothesis'' naturally gives rise to the Max-Min: when statistically similar chain-like systems are compared -- either by an evolutionary process or by industrial quality testing -- the system with the strongest weakest link prevails.    

The Min-Max also arises naturally from real-world applications. Indeed, consider a back-up system in which critical files are stored on multiple separate hard drives. If a file is damaged on one of the drives it could be retrieved from another; however, if all copies of a file are damaged then the file is lost forever. The loss time of a given file is thus the maximum of its damage times over the different drives. In turn, since all files are critical, system failure occurs at the first loss time of a file. Thus, the system failure time is the Min-Max of the files' damage times.

Here we adopted the setting of random-matrix theory, considering large matrices with IID entries. For the Max-Min and Min-Max of such matrices we established, respectively, the Gumbel approximations of Eqs. (\ref{21}) and (\ref{22}); also, we showed how to apply these approximations as a `scientific  tool' and as an `engineering tool'. The approximations stem from the limit-laws of Eqs. (\ref{11}) and (\ref{12}) -- which assume the role of a ``Gumbel Central Limit Theorem'' for the Max-Min and Min-Max. With their generality and universality, their easy practical implementation, and their many potential applications -- e.g. in game theory, in reliability engineering, in materials science, and in the design of back-up systems -- the novel results presented herein are expected to serve diverse audiences in the physical sciences and beyond.

\textbf{Acknowledgments}. R.M. acknowledges Deutsche Forschungsgemeinschaft
for funding (ME 1535/7-1) and support from the Foundation for Polish Science
within an Alexander von Humboldt Polish Honorary Research Fellowship. S.R.
gratefully acknowledges support from the Azrieli Foundation and the Sackler
Center for Computational Molecular and Materials Science.

\end{document}